\documentstyle[psfrag,aps,preprint,epsfig,axodraw]{revtex}

\begin{document}

\tighten

\preprint{TU-714}
\title{Large Supersymmetric Contribution to CP Asymmetry of  
$B_d \to \phi K_S$ 
from Left-Handed Squark Mixing}
\author{Motoi Endo, Mitsuru Kakizaki and Masahiro Yamaguchi}
\address{ Department of Physics, Tohoku University,
Sendai 980-8578, Japan}
\date{\today}
\maketitle
\begin{abstract}
  Supersymmetric contribution to the CP asymmetry of $B_d \to \phi
  K_S$ is re-examined.  It is emphasized that our analysis takes into
  account the recently found constraint from the mercury electric
  dipole moment, in addition to the well-known constraint from the
  branching ratio of $b \to s \gamma$. We show that, despite these
  constraints, the CP asymmetry can considerably deviate from the
  Standard Model prediction, if the CP-odd flavor mixing comes from
  left-handed squarks.  Alignment mechanism of sfermion masses and
  mixing may imprint the required flavor mixing at high energy scale.
\end{abstract}

\clearpage
\section{Introduction}

One of the most celebrated features of the Standard Model of particle
physics is that the flavor changing neutral current (FCNC) is suppressed by
the GIM mechanism.  All contributions to the FCNC processes arise at loop
level, which are further suppressed by either small mass differences between
different generations or small generation mixing angles. In fact, the mixing
between the first two generations is suppressed because of the small quark
masses (and thus small mass differences) and the mixing containing the third
generation is suppressed because of the small entries of the 
Cabibbo-Kobayashi-Maskawa (CKM) matrix.

Extensions of the Standard Model generically do not have such a
suppression mechanism and thus tend to suffer  too large FCNC.
This fact gives stringent constraints on how new physics beyond the
Standard Model should be. Put another way, it implies that signals of
a beyond-the-Standard-Model may be revealed in flavor physics.

Nowadays a special attention is  paid to the
flavor mixings in B mesons. On the experimental side, B-factories are
in operation, with a lot of new results on the nature of the flavor
mixings of the B mesons. On the theoretical side, the fact that the
quark masses are hierarchical in inter-generations implies that the
third generation may be special and a new physics may manifest itself
there.

Among other things, a very interesting process is $B_d \to \phi K_S$
whose CP asymmetry  is measured at the B factories. Theoretically a new physics
contribution may easily be seen in this process because there is no
tree-level contribution in the Standard Model. The latest
result of the Belle collaboration announced that the discrepancy from
the Standard Model in the CP asymmetry of $B_d \to \phi K_S$ is still
there \cite{Abe:2003yt}, {\it i.e.}
\begin{eqnarray}
  S_{\phi K_S} = -0.96 \pm 0.50^{+0.09}_{-0.11},
\end{eqnarray}
while that of the BaBar collaboration \cite{Aubert:2004ii}
\begin{eqnarray}
  S_{\phi K_S} = 0.47 \pm 0.34^{+0.08}_{-0.06}
\end{eqnarray}
is consistent with the
Standard Model prediction. Though the
situation is yet unclear, one certainly has to watch what is happening
when more data are accumulated, and on the theoretical side it is
important to investigate in what situation new physics can generate large $b
\to s$ transition.

Here we consider supersymmetric models, a promising framework to solve
the gauge hierarchy problem. A new source of flavor mixing originates
from SUSY breaking masses of squarks and sleptons.  The flavor mixing between
the second and third generations both in
left-handed squarks (LL) and  in right-handed squarks (RR) can
make contribution to $B_d \to \phi K_S$. Combined with flavor-conserving 
left-right squark mixing  which
is proportional to $m_b \tan \beta$, with
$\tan \beta$ being the ratio of the vacuum expectation values of the two
Higgs doublets, the flavor mixing in the left-right squark mixing can also 
be induced.\footnote{Throughout this paper, we do not
  consider the direct flavor transition in the left-right squark
  mixing.} 

The flavor mixing in the RR sector can be generated by renormalization
group effects of right-handed neutrino Yukawa couplings if the theory
is grand unified. The implications to the CP asymmetry of $B \to \phi K_s$
and other processes were studied 
intensively~\cite{Moroi:2000tk,Chang:2002mq,Harnik:2002vs}. More
general analyses were also done in 
Refs.~\cite{Khalil:2002fm,Kane:2002sp,Ciuchini:2002uv,Baek:2003kb,Goto:2003iu,Arnowitt:2003ev,Ciuchini:2003rg}.

A notable observation has recently been made in Ref. \cite{Hisano:2003iw},
which has found that the present experimental bound on the electric
dipole moment (EDM) of mercury atom severely constrains the CP
violating part of a product of the LL and RR down-type squark flavor
mixing between the 2nd and the 3rd generations.  Radiative corrections 
(renormalization group effects) due to
Yukawa interaction for up-type quark masses generate significant flavor
mixing in the LL sector when one considers the high scale SUSY breaking
scenario where the mediation of the SUSY breaking takes place at a
high energy scale. Given the non-negligible mixing in the LL sector at
the electroweak scale, the bound from the $^{199}$Hg EDM thus practically
constrains the flavor mixing in the RR sector.  It has been shown that
with the parameters satisfying the constraint the contribution to the
CP asymmetry of the $B_d \to \phi K_S$ decay from the RR mixing is
negligibly small.

We are thus led to consider the LL mixing in the squark masses as a
possible source of the beauty to strange transition in the SUSY
models. In a recent paper \cite{Endo:2003te}, we have discussed 
the constraint on the LL flavor mixing coming from the Hg EDM.  
In this case the dominant contribution is from
chargino-squark loop and thus it is not as large as the gluino-squark
loop contribution. Thus one expects that the SUSY contribution to the
CP asymmetry of the $B_d \to \phi K_S$ decay can be considerably large
even when presently known constraints including that from the
Hg EDM are taken into account.

A purpose of the present paper is to quantify this prospect.  We will
exhibit our numerical study on the CP asymmetry of $B_d \to \phi K_S$.
We emphasize that the new constraint from the Hg EDM will be taken into 
account, in addition to the well-known constraint from $b \to s \gamma$.
 We will show that
the SUSY contribution to the CP asymmetry can, in fact, be large when
$\tan \beta$ is large even
after the constraints are imposed.   
We will also observe that the constraint from the Hg EDM is in general 
comparable to that of $b \to s \gamma$ in severity, and which of
the two is more stringent depends on the parameters chosen.  This
illustrates the importance of the new constraint from the Hg EDM
discussed in our previous paper.

\section{Flavor Structure of Squark Masses}

Before going to the flavor changing processes, we would like to
briefly state the flavor structure of the squark mass matrices assumed
in this paper.  The soft SUSY breaking sfermion masses are given at a very high energy
scale, which are driven down to low energy by renormalization group
flow.  The flavor violation in the sfermion masses thus has two
sources: one is imprinted at the high energy scale, and the other one
is due to renormalization group effects.  In the minimal supergravity,
there is no imprinted flavor violation in the sfermion masses, and the
flavor mixing comes only from that of the Yukawa coupling matrices.
In the minimal supersymmetric model, the flavor mixing arises solely
in the SU(2)$_L$ doublet (left-handed) squark mass matrix (LL sector) 
and CP phase
is proportional to the Kobayashi-Maskawa CP phase in the quark Yukawa
coupling matrix. Thus in this case the SUSY contribution to the CP
asymmetry of $B_d \to \phi K_S$ is not expected. On the other hand, in
a grand unified theory with right-handed neutrino Yukawa couplings,
the SU(2)$_L$ singlet (right-handed) down-type squarks (RR sector) have flavor
mixing as well as new CP violation. It was argued that this gives a
sizable SUSY contribution to the CP asymmetry of $B_d \to \phi
K_S$~\cite{Moroi:2000tk,Chang:2002mq,Harnik:2002vs}. However the CP phase would make too large a contribution to the
Hg EDM unless there is unplausible cancellation between different diagrams
\cite{Hisano:2003iw}.

Our hypothetical flavor-mixing with new CP phase in the SU(2)$_L$
doublet (left-handed) squarks, namely in the LL sector, is likely due
to the imprinting at the ultra high energy scale, where the flavor
violation of the squark masses appears from the beginning.  Namely we
assume that the initial soft SUSY breaking masses are given such that
flavor violation arises only in the LL sector. Furthermore we assume
that the renormalization group effect which causes flavor mixing
solely comes from the CKM matrix. The latter implies that, in the
context of GUT, either the right-handed neutrino Yukawa couplings are
small or the soft masses arise below the GUT scale.  We anticipate 
existence of alignment
mechanism of quark/squark masses and mixing which realizes the
required squark flavor mixing pattern, possibly due to flavor symmetry
or geometry of extra dimensions. Actually the democratic sfermion
mechanism advocated in Ref. \cite{Hamaguchi:2002vi} can give this mass
pattern: the flavor mixing in the right-handed down-type squarks,
which form $\bar 5$ multiplet with left-handed (s)leptons, is argued
to be absent due to non-trivial interplay with neutrino mass
matrix. Here we will not pursue possible alignment mechanisms to
realize this structure of the squark mass matrices any further in this
paper. Rather we simply assume this structure and consider its
phenomenological implications.

\section{SUSY contribution to $B_d \to \phi K_S$}

Let us turn to  SUSY contributions to $B_d \to \phi K_S$.  The
time dependent CP asymmetry of the $B_d \to \phi K_S$ is defined as
\begin{eqnarray}
  a_{\phi K}(t) = C_{\phi K} \cos(\Delta M_{B_d} t) 
                + S_{\phi K} \sin(\Delta M_{B_d} t),
\end{eqnarray}
where $C_{\phi K}$ and $S_{\phi K}$ are given by
\begin{eqnarray}
  C_{\phi K} = \frac{1-|\lambda|^2}{1+|\lambda|^2},~~~
  S_{\phi K} = \frac{2 {\rm Im}\lambda}{1+|\lambda|^2}.
\end{eqnarray}
Here $\lambda$ is defined as
\begin{eqnarray}
  \lambda = \frac{q}{p} 
  \frac{\bar{{\mathcal A}}(\bar{B}_d \to \phi K_S)}
       {{\mathcal A}(B_d \to \phi K_S)}
  = \frac{q}{p}
  \frac{(\bar{\mathcal A}_{\phi K}^{\rm SM} + 
    \bar{\mathcal A}_{\phi K}^{\rm SUSY})}
  {({\mathcal A}_{\phi K}^{\rm SM} + {\mathcal A}_{\phi K}^{\rm SUSY})}.
\end{eqnarray}
The ratio $q/p$ is evaluated from the $B_d^0 - \bar{B}_d^0$ mixing and is 
approximately parameterized as $q/p = e^{2i\beta}$.  If the decay amplitude 
is almost determined by the SM diagrams and there is no additional CP 
violation, the ratio $\bar{\mathcal A}/{\mathcal A}$ is real and the CP 
asymmetry $S_{\phi K}$ becomes the same as the one measured by the 
$B_d \to J/\psi K_S$ process, $S_{J/\psi K} = \sin 2\beta = 0.736 \pm 0.049$
\cite{BelleCP,BabarCP}.  
The SUSY contributions to $B_d \to \phi K_S$ in the decay amplitudes is, 
however, generically comparable to the SM one because the latter is 
loop suppressed.  Then the CP asymmetry $S_{\phi K}$ deviates from the value 
by $B_d \to J/\psi K_S$.  

The $\Delta B = 1$ effective Hamiltonian which is relevant for 
the $b \to s$ transition is given by
\begin{eqnarray}
  {\mathcal H}_{\rm eff} = -\frac{4G_F}{\sqrt{2}} V_{tb}V_{ts}^* 
  \left[
    \sum_{i=2}^6 (C_i O_i + C'_i O'_i) + 
    C_{7\gamma} O_{7\gamma} + C'_{7\gamma} O'_{7\gamma} + 
    C_{8g} O_{8g} + C'_{8g} O'_{8g}
  \right].
\end{eqnarray}
where
\begin{eqnarray}
  O_2 &=& (\bar{s} \gamma^\mu P_L c) (\bar{c} \gamma_\mu P_L b),
  \nonumber \\
  O_3 &=& (\bar{s} \gamma^\mu P_L b) (\bar{s} \gamma_\mu P_L s),
  \nonumber \\
  O_4 &=& (\bar{s}_i \gamma^\mu P_L b_j) (\bar{s}_j \gamma_\mu P_L s_i),
  \nonumber \\
  O_5 &=& (\bar{s} \gamma^\mu P_L b) (\bar{s} \gamma_\mu P_R s),
  \\
  O_6 &=& (\bar{s}_i \gamma^\mu P_L b_j) (\bar{s}_j \gamma_\mu P_R s_i),
  \nonumber \\
  O_{7\gamma} &=& 
  \frac{e}{16\pi^2}m_b(\bar{s}_i\sigma^{\mu\nu}P_Rb_i)F_{\mu\nu},
  \nonumber \\
  O_{8g} &=& 
  \frac{g_s}{16\pi^2}m_b(\bar{s}_i\sigma^{\mu\nu}T_{ij}^aP_Rb_j)G_{\mu\nu}^a.
  \nonumber
\end{eqnarray}
Here $i$ and $j$ are color indices, $P_{R,L} = (1\pm\gamma_5)/2$.  
The prime expresses the exchange of $L$ and $R$.

We have 5 four-quark operators and 2 dipole ones for the flavor changing 
process.  The corresponding Wilson coefficients $C_i$ are evaluated from the 
diagrams in which the the heavy SM or the SUSY particles propagate virtually.  
In the low energy region, these coefficients have two sources of flavor 
violation: the CKM matrix and squark mass matrices.  
The squark flavor mixings are denoted by the following 
squark flavor mixing parameters in the mass insertion approximation (MIA)
\begin{eqnarray}
  & & (\delta^u_{LL})_{ij}=\frac{(m^2_{\tilde u_L})_{ij}}{m^2_{\tilde q}}, 
  \quad
  (\delta^u_{RR})_{ij}=\frac{(m^2_{\tilde u_R})_{ij}}{m^2_{\tilde q}},
  \nonumber \\
  & & (\delta^u_{LR})_{ij} 
  = \frac{(m^2_{\tilde uLR})_{ij}}{m^2_{\tilde{q}}}, \quad
  \quad (\delta^u_{RL})_{ij} = (\delta^{u*}_{LR})_{ji}.
\end{eqnarray}

First, we consider the flavor mixing whose origin comes from the squark 
mass terms.  Since the RR mixing should be strongly suppressed considering 
the Hg EDM \cite{Hisano:2003iw}, we concentrate on the LL squark mass matrix.  
In MIA, we estimate the Wilson coefficients at the SUSY mass scale at the 
one loop level as 
\begin{eqnarray}
  C_3({\rm SUSY}) &=& 
  \frac{\sqrt{2}\alpha_s^2}{4G_FV_{tb}V_{ts}^*m_{\tilde{q}}^2}
  (\delta_{LL}^d)_{23}\left[
    -\frac{1}{9}B_1(x)-\frac{5}{9}B_2(x)-\frac{1}{18}P_1(x)-\frac{1}{2}P_2(x)
  \right],
  \nonumber \\
  C_4({\rm SUSY}) &=& 
  \frac{\sqrt{2}\alpha_s^2}{4G_FV_{tb}V_{ts}^*m_{\tilde{q}}^2}
  (\delta_{LL}^d)_{23}\left[
    -\frac{7}{3}B_1(x)+\frac{1}{3}B_2(x)+\frac{1}{6}P_1(x)+\frac{3}{2}P_2(x)
  \right],
  \nonumber \\
  C_5({\rm SUSY}) &=& 
  \frac{\sqrt{2}\alpha_s^2}{4G_FV_{tb}V_{ts}^*m_{\tilde{q}}^2}
  (\delta_{LL}^d)_{23}\left[
    \frac{10}{9}B_1(x)+\frac{1}{18}B_2(x)-\frac{1}{18}P_1(x)-\frac{1}{2}P_2(x)
  \right],
  \nonumber \\
  C_6({\rm SUSY}) &=& 
  \frac{\sqrt{2}\alpha_s^2}{4G_FV_{tb}V_{ts}^*m_{\tilde{q}}^2}
  (\delta_{LL}^d)_{23}\left[
    -\frac{2}{3}B_1(x)+\frac{7}{6}B_2(x)+\frac{1}{6}P_1(x)+\frac{3}{2}P_2(x)
  \right],
  \nonumber \\
  C_{7\gamma}({\rm SUSY}) &=& 
  -\frac{\alpha_s\pi}{3\sqrt{2}G_FV_{tb}V_{ts}^*m_{\tilde{q}}^2}
  (\delta_{LL}^d)_{23} \left[
    \frac{8}{3}M_3(x) - 
    \mu_H \tan\beta \frac{m_{\tilde{g}}}{m_{\tilde{q}}^2}\frac{8}{3}M_a(x)
  \right],
  \nonumber \\
  C_{8g}({\rm SUSY}) &=& 
  \frac{\alpha_s\pi}{\sqrt{2}G_FV_{tb}V_{ts}^*m_{\tilde{q}}^2}
  (\delta_{LL}^d)_{23} \left[
    \frac{8}{3}\left(-\frac{1}{3}M_3(x) - 3M_4(x)\right) 
  \right.
  \nonumber \\
  &&
  \left.
    - \mu_H \tan\beta \frac{m_{\tilde{g}}}{m_{\tilde{q}}^2}
    \left(-\frac{1}{3}M_a(x)-3M_b(x)\right)
  \right],
  \label{eq:WilsonCoefficients}
\end{eqnarray}
where $B(x)$, $P(x)$ and $M(x)$ are the loop functions 
\footnote{The loop functions $B(x)$, $P(x)$, $M_3(x)$ and $M_4(x)$ are 
defined in Ref. \cite{Gabbiani:1996hi}.  $M_a(x)$ and $M_b(x)$ are same as 
$M_1(x)$ and $M_2(x)$ of Ref. \cite{Hisano:2003iw}, respectively.  
} and $x = m_{\tilde{g}}^2/m_{\tilde{q}}^2$.  
And $\mu_H$, $m_{\tilde{g}}$ and $m_{\tilde{q}}$ are the higgsino mass 
parameter, the gluino mass and typical squark mass, respectively.  
Here we consider the gluino mediated diagrams which dominate the SUSY 
contributions.  
We note here that some terms in Eq. (\ref{eq:WilsonCoefficients}) are 
enhanced by $\tan\beta$.  These are due to double mass insertion diagrams, 
which contain both the (flavor violating) LL mixing and the (flavor 
conserving) RL down type squark mixing parameters.  
This type of enhancement is dominant for the Wilson 
coefficients which require chirality flipping, that is, $C_{7\gamma}$ 
and $C_{8g}$.  

The other contributions stem from the CKM matrix.  These include the SM, 
the charged Higgs, the chargino and the neutralino diagrams.  
The SM ones are estimated at the one loop order and partially at the two 
loop level \cite{Buchalla:1995vs}.  And the charged Higgs diagrams are 
calculated at the two loop order \cite{Ciuchini:1997xe}.  
The others from the SUSY particles are estimated at the 
one loop level, including the corrections by large $\tan\beta$ 
\cite{Degrassi:2000qf}, \cite{Borzumati:2003rr}.  

We can estimate the decay amplitude of $B_d \to \phi K_S$ in terms of these 
Wilson coefficients.  
The SUSY contributions to the CP asymmetry of $B_d \to \phi K_S$ are dominated 
by the chromo-magnetic operator, and thus enhanced by $\tan \beta$.  
Here we use the naive factorization ansatz.  The matrix element of 
chromo-magnetic moment is given by
\begin{eqnarray}
  \langle \phi K_S|
  \frac{g_s}{16\pi^2}m_b(\bar{s}_i\sigma^{\mu\nu}T_{ij}^aP_Rb_j)G_{\mu\nu}^a
  |\bar{B}_d\rangle = 
  \kappa \frac{2\alpha_s}{9\pi}(\epsilon_\phi p_B)
  f_\phi m_\phi^2 F_+(m_\phi^2),
\end{eqnarray}
where $\epsilon_\phi$, $m_\phi$ and $f_\phi$ are the polarization vector, 
the mass and the decay constant of $\phi$ meson, respectively.  
And $F_+(m_\phi^2)$ is the B to K transition form factor.  
The coefficient $\kappa = -1.1$ is estimated in the heavy-quark effective 
theory \cite{Harnik:2002vs}, though there is large hadronic uncertainty.  
Then the SUSY contributions to the $B_d \to \phi K_S$ decay amplitude 
depend on the LL squark flavor mixing parameter, the squark mass, 
$\tan\beta$ and $\mu_H$.  

The squark flavor mixing $(\delta_{LL}^d)_{23}$ is constrained by the 
branching ratio of $b \to s \gamma$ and the Hg EDM.  The experimental value 
of the $b \to s \gamma$ branching ratio well agrees with the prediction by 
the SM.  Thus the SUSY contributions have to be canceled among them.  
Experimentally, the branching ratio $Br(\bar{B} \to X_s \gamma)$ was measured 
precisely 
\cite{Barate:1998vz,Chen:2001fj,Abe:2001hk,Aubert:2002pb,jessop}.
Here we take 
\begin{eqnarray}
  2.0 \times 10^{-4} < Br(b \to s \gamma) < 4.5 \times 10^{-4},
\end{eqnarray}
which is rather conservative in the light of various theoretical 
uncertainties.  
In the absence of the squark mixing, the amplitude from the chargino 
diagrams is generically comparable to those from the SM.  
A proper choice of sign of $\mu_H$ causes the destructive interference among 
the SM and the SUSY diagrams.  On the other hand, the contributions from the 
imaginary part of the squark flavor mixing is always additive to the branching 
ratio.  The electromagnetic moment in Eq. (\ref{eq:WilsonCoefficients}) 
dominates the branching ratio.  Then the relevant parameters are 
$(\delta_{LL}^d)_{23}$, $\tan\beta$, $\mu_H$ and the soft masses 
$m_{\tilde{g}}^2$ and $m_{\tilde{q}}^2$.  

The Hg EDM also gives a bound on the squark flavor mixing parameter.  
In fact, although the uncertainty comes from the hadron dynamics, 
the Hg EDM constrains the imaginary part of up type squark flavor mixing 
$(\delta_{LL}^u)_{23}$ generically.  This is because the strange quark EDM 
affects the Hg EDM.  The flavor mixing causes the EDM via the chargino loop.  
This diagram is also enhanced by $\tan\beta$.  
Thus the EDM is enhanced by the large LL squark flavor mixing and/or large 
$\tan \beta$, and decreases when squark mass is large.  
The Hg EDM also depends on the trilinear coupling $A_t$ through the LR mixing 
rather than $\mu$ parameter, which appears in the diagram only as the charged 
higgsino mass and mixing.  From the existing experimental data,
$|d_{Hg}|<2.1 \times 10^{-28}\ e\ \mbox{cm}$ at 95\% C.L. 
\cite{Romalis:2000mg}, we obtain the constraint on the CEDM of the 
strange quark \cite{Falk:1999tm},
\begin{equation}
    e|d_s^C| < 5.5 \times 10^{-25}\ e\ \mbox{cm},
\end{equation}
where the definition of $d_s^C$ is given in Ref. \cite{Endo:2003te}.  
In the generic situation, this bound on the left-handed up type 
squark mass is applied for the left-handed down type squark one, 
\begin{equation}
  (\delta^d_{LL})_{32} \simeq (\delta^u_{LL})_{32} 
  + \lambda (\delta^u_{LL})_{31} + O(\lambda^2),
\end{equation}
with $\lambda \simeq 0.22$.
Hereafter we consider the generic situation and assume $(\delta^d_{LL})_{32} 
\simeq (\delta^u_{LL})_{32}$.  
Then we can constrain the decay amplitude of the $B_d \to \phi K_S$.  
In particular, this bound from the Hg EDM becomes severer than that from 
$b \to s \gamma$ when $\mu_H$ is small.  

\begin{figure}[h]
  \begin{center}
    \includegraphics[scale=0.8]{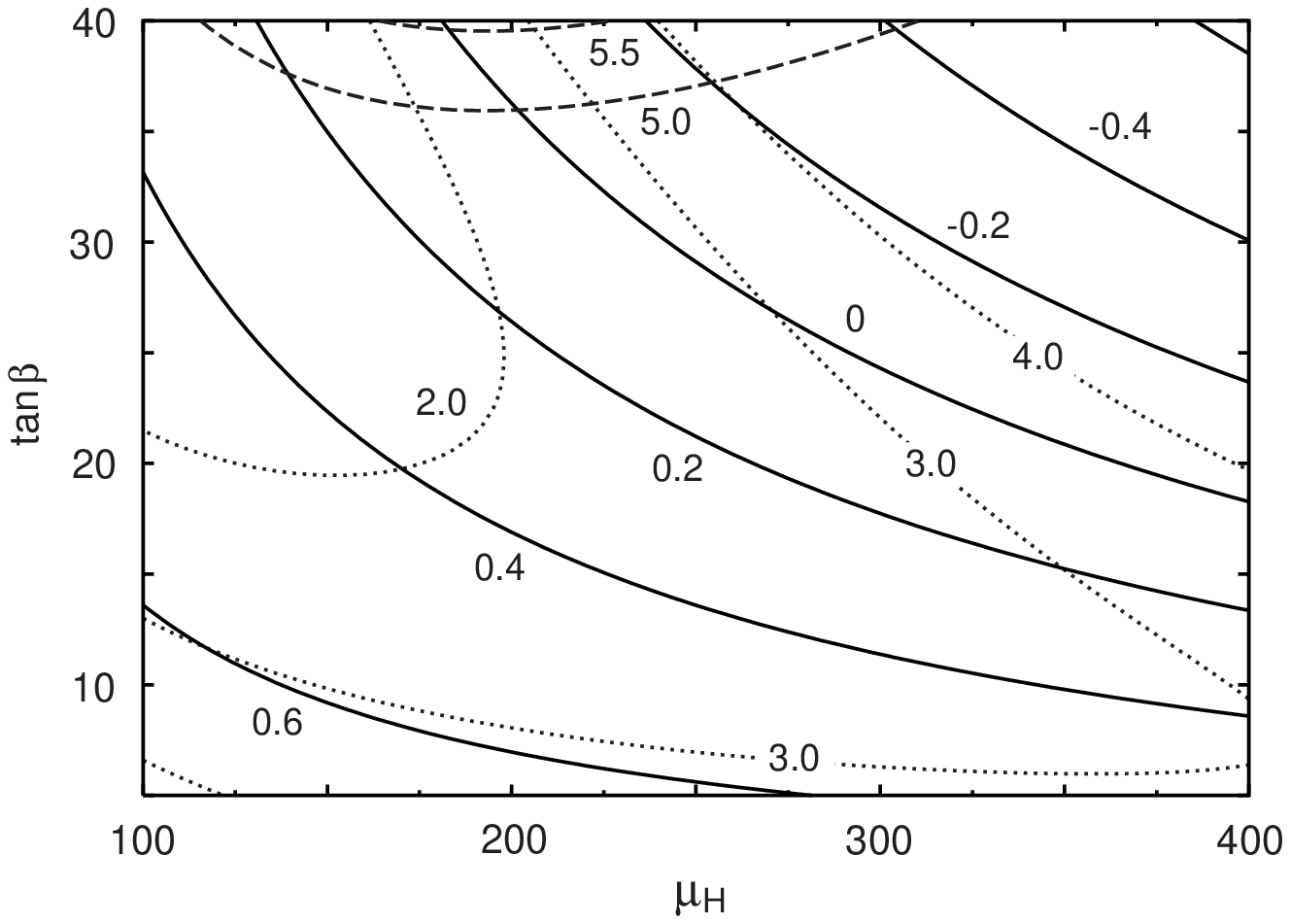}
  \end{center}
  \caption{Constant contours of the CP asymmetry $S_{\phi K_S}$ (solid) 
    when the LL down-type squark mixing is $|(d_{LL}^d)_{23}|=0.1$ with 
    maximal phase.  The contours of the $b \to s \gamma$ branching ratio 
    (dotted, in units of $10^{-4}$) and the CEDM of the strange quark 
    which is constrained from the Hg EDM (dashed, 5.0 and 5.5 in 
    units of $10^{-25}$) are also shown.  The soft parameters are 
    $m_{\tilde{g}}^2 = m_{\tilde{q}}^2 = (500\ {\rm GeV})^2$ and 
    $A_t = 500\ {\rm GeV}$.}
  \label{fig:d01}
\end{figure}

\begin{figure}[h]
  \begin{center}
    \includegraphics[scale=0.8]{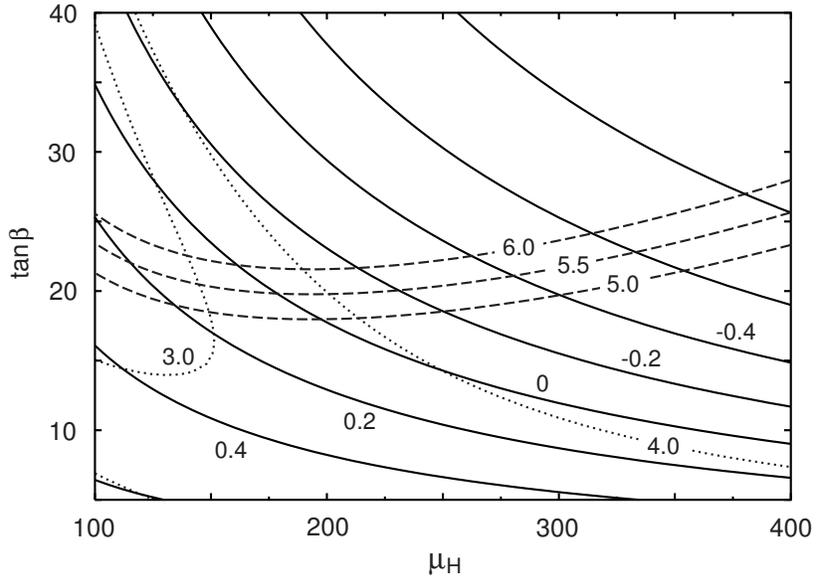}
  \end{center}
  \caption{Same as Fig. \ref{fig:d01} but $|(d_{LL}^d)_{23}|=0.2$.}
  \label{fig:d02}
\end{figure}

With the experimental constraints given above, we study the CP asymmetry 
of $B_d \to \phi K_S$ numerically.  In Fig. \ref{fig:d01} and \ref{fig:d02}, 
the constant contours of the CP asymmetry of $B_d \to \phi K_S$ are shown.  
The constraints from the $b \to s \gamma$ branching ratio and the Hg EDM 
are also displayed on the graph.  
These quantities depends mainly on $(\delta_{LL}^u)_{23}$, $\tan\beta$ 
, $\mu_H$ (and partially on $A_t$) and the soft masses $m_{\tilde{g}}^2$ and 
$m_{\tilde{q}}^2$.  
Here we show the result of $|(\delta_{LL}^u)_{23}| = 0.1$ and $0.2$ 
with maximal complex phase.  
As a reference, we also fix the soft parameters as $m_{\tilde{g}}^2 = 
m_{\tilde{q}}^2 = (500\ {\rm GeV})^2$ and $A_t = 500\ {\rm GeV}$.  
And we take the other model parameters: the Wino mass $M_2 = 250\ {\rm GeV}$ 
and the Higgs mass parameters $m_{Hd}^2 = -m_{Hu}^2 = (250\ {\rm GeV})^2$.  
The CP asymmetry becomes large as $\mu_H$ or $\tan\beta$ increases.  
From Fig. \ref{fig:d01} and \ref{fig:d02}, we perceive that the contribution 
from the LL mixing to $B_d \to \phi K_S$ can become as large as 
$S_{\phi K} \lesssim 0$.  

We find that the resultant maximal difference between $S_{\phi K_S}$ and 
$S_{J/\psi K_S}$ is insensitive to $(\delta_{LL}^u)_{23}$.  
This is because $S_{\phi K_S}$ and both constraints are all controlled 
by the mixing parameter.  The value also depends weakly on the soft masses, 
$M_2$ and $m_H$.    We also note that since the SUSY contribution to 
the Hg EDM is proportional to $A_t$, larger $A_t$ reduces the possible 
maximal value of deviation of $S_{\phi K_S}$ gradually 
(See Fig.\ref{fig:d01} and \ref{fig:d02}).  
Thus we conclude that the CP asymmetry of $B_d \to \phi K_S$ can deviate 
sizably from that of $B_d \to J/\psi K_S$ by the LL squark flavor mixing 
with the RR mixing suppressed.  

Here we would like to briefly mention implications of the LL squark
flavor mixing to other $b \to s$ transition processes.  First let us
consider $B_s^0 - \bar{B}_s^0$ mixing.  The dominant diagram of the
SUSY contribution is not enhanced by $\tan\beta$ and $\mu_H$ and
depends only on the LL squark flavor mixing parameter.  Thus the
effect from SUSY particles is expected to be small.  For
$|(\delta_{LL}^u)_{23}| = 0.2$ and
$m_{\tilde{g}}^2=m_{\tilde{q}}^2=(500\ {\rm GeV})^2$, the SUSY
contribution to $B_s$ mixing is about $\Delta M_{B_s}|_{\rm SUSY}
\simeq 2 {\rm ps}^{-1}$, which may be too small to de identified as a
signal of new physics at Tevatron Run II.  
We have also checked the effect on the
CP asymmetry of $b \to s \gamma$. We expect the CP asymmetry to be at most 
8\%, which may marginally be seen at the on-going experiments.

\section{Summary}
In this paper, we revisited the SUSY contribution to the CP asymmetry
of $B_d \to \phi K_S$, in the light of the recently found constraint
from the mercury EDM. Ref.~\cite{Hisano:2003iw} had pointed out that
the flavor mixing of the squarks in the RR sector receives a very
severe constraint from the mercury EDM, provided that the 2nd and 3rd
generation mixing in the LL sector is generated non-zero by
renormalization group effect from the quark Yukawa couplings. Thus the
flavor mixing in the RR sector will not give significant contribution
to the $B_d \to \phi K_S$ CP asymmetry, unless the constraint from the
mercury EDM is ameliorated by accidental cancellation. This argument
allows for the flavor mixing in the LL sector to be practically the
only source to generate sizable effect to the CP asymmetry. In our
recent paper \cite{Endo:2003te}, we pointed out that the 2nd-3rd
generation mixing in the LL sector is still constrained by the mercury
EDM, but much less than the previous case. Combined with the
well-known constraint from $Br(b \to s \gamma)$, we identified the
allowed region of the parameter space, and showed that, in the allowed
region, the CP asymmetry to $B_d \to \phi K_S$ can sizably deviate
from the Standard Model prediction.

The origin of the flavor mixing in the LL sector is speculated as the
alignment mechanism of quark/squark mass matrices. It is interesting
to point out that the democratic sfermion masses considered in
\cite{Hamaguchi:2002vi} can provide the desired flavor structure for
the squarks. Attempts to seek for other possible alignment mechanisms based
on flavor symmetry or geometry of extra dimensions should be encouraged.

\section*{Acknowledgment}
This work was supported in part by the Grants-in-aid from the Ministry
of Education, Culture, Sports, Science and Technology, Japan, No.12047201 and
No.14046201.
ME and MK thank the Japan Society for the Promotion of Science for financial 
support.


\end{document}